\documentclass[twocolumn,superscriptaddress,showpacs,preprintnumbers,amsmath,amssymb,prb]{revtex4}

\usepackage{graphicx}
\usepackage{dcolumn}
\usepackage{bm}
\usepackage{epsf}

\usepackage{color}
\definecolor{gold}{rgb}{0.85,.66,0}

\begin{document}

\preprint{APS/123-QED}

\title{Dynamics of exciton magnetic polarons in CdMnSe/CdMgSe quantum wells: \\ the effect of self-localization}

\author{I.~A.~Akimov}
 \affiliation{Experimentelle Physik 2, Technische Universit\"at Dortmund, 44221 Dortmund, Germany}
 \affiliation{Ioffe Physical-Technical Institute, Russian Academy of Sciences, 194021 St. Petersburg, Russia }
\author{T.~Godde}
 \affiliation{Department of Physics and Astronomy, University of Sheffield, Sheffield S3 7RH, United Kingdom}
\author{K.~V.~Kavokin}
\affiliation{Ioffe Physical-Technical Institute, Russian Academy of Sciences, 194021 St. Petersburg, Russia }
\author{D.~R.~Yakovlev}
 \affiliation{Experimentelle Physik 2, Technische Universit\"at Dortmund, 44221 Dortmund, Germany}
 \affiliation{Ioffe Physical-Technical Institute, Russian Academy of Sciences, 194021 St. Petersburg, Russia }
\author{I.~I.~Reshina}
\author{I.~V.~Sedova}
\author{S.~V.~Sorokin}
\author{S.~V.~Ivanov}
\author{Yu.~G.~Kusrayev}
 \affiliation{Ioffe Physical-Technical Institute, Russian Academy of Sciences, 194021 St. Petersburg, Russia }

\author{M.~Bayer}
 \affiliation{Experimentelle Physik 2, Technische Universit\"at Dortmund, 44221 Dortmund, Germany}
 \affiliation{Ioffe Physical-Technical Institute, Russian Academy of Sciences, 194021 St. Petersburg, Russia }

\date{\today}

\begin{abstract}
We study the exciton magnetic polaron (EMP) formation in (Cd,Mn)Se/(Cd,Mg)Se diluted-magnetic-semiconductor quantum wells using time-resolved photoluminescence (PL). The magnetic field and temperature dependencies of this dynamics allow us to separate the non-magnetic and magnetic contributions to the exciton localization. We deduce the EMP energy of 14~meV, which is in agreement with time-integrated measurements based on selective excitation and the magnetic field dependence of the PL circular polarization degree. The polaron formation time of 500~ps is significantly longer than the corresponding values reported earlier. We propose that this behavior is related to strong self-localization of the EMP, accompanied with a squeezing of the heavy-hole envelope wavefunction. This conclusion is also supported by the decrease of the exciton lifetime from 600~ps to 200 - 400~ps with increasing magnetic field and temperature.
\end{abstract}

\maketitle

\section{Introduction}
\label{sec:Introduction}

In diluted magnetic semiconductors (DMS) carriers are coupled with the localized spins of the magnetic ions by strong exchange interaction \cite{Furdyna, Dietl}. This leads to giant magneto-optical effects that have been investigated in depth for II-VI DMS with magnetic Mn$^{2+}$ ions, like (Cd,Mn)Te, (Cd,Mn)Se and (Zn,Mn)Se and their heterostructures \cite{DMS_book, Dietl, Spinbook}. The giant Zeeman splitting of the band states, the giant Faraday and Kerr rotation and the existence of magnetic polaron are among the most widely studied phenomena.

The magnetic polaron (MP) formation is caused by ferromagnetic alignment of the localized magnetic ions in the vicinity of a carrier. Carrier localization plays a crucial role for the MP stability and controls the polaron energy. Carriers may be bound to impurity centers (donors or acceptors) or localized by electrostatic or magnetic potential fluctuations, like alloy fluctuations in ternary semiconductors or well width fluctuations in quantum wells. As a result, bound or localized magnetic polarons can be studied in optical spectra, see e.g. Refs.~\onlinecite{DMS_book2, Dietl, Number6, Yakovlev10}. Also criteria for the stability of free magnetic polarons, i.e. for magnetic polaron formation without initial localization of the involved carrier, have been considered theoretically for II-VI semiconductors~\cite{Ben93,Theory}. In particular, it has been shown that free magnetic polarons are not stable in three dimensional systems so that their formation cannot be expected in bulk samples. However, reduction of the system's dimensionality to two favors free MP stability.

These predictions stimulated experimental studies of exciton magnetic polarons (EMP) in quantum well (QW) structures based on (Cd,Mn)Te and (Zn,Mn)Se~\cite{SSC90, SSC93,Yakovlev92, Yakovlev95, Poweleit94, Henneberger96}. The reduction of dimensionality provided by decreasing the QW width indeed favors EMP formation and increases its binding energy, which can reach values up to 30~meV. However, truly free MPs have not been found experimentally so far, only partial self-localization during MP formation has been established from studying the MP dynamics~\cite{Number6, Linz94}. The properties of two-dimensional EMPs have been studied in great detail for (Cd,Mn)Te/(Cd,Mg)Te heterostructures. The dependencies of the polaron energy on the structure design and experimental conditions, like external magnetic field strength and temperature were measured by means of optical selective excitation~\cite{Yakovlev10, Kusrayev-89, Mackh94,Mackh94b,Kusrayev-96}. The spin dynamics of carriers and localized magnetic ions during the polaron formation process were studied by time-resolved spectroscopy and polarization sensitive techniques. Based on the current understanding it is important to extend these investigations to other material systems in order to check whether the conclusions drawn for EMPs in (Cd,Mn)Te/(Cd,Mg)Te heterostructures can be generalized, and also to search for new EMP regimes which may be offered by novel DMS systems.

In this paper we present the results of optical studies of exciton magnetic polarons in (Cd,Mn)Se/(Cd,Mg)Se DMS heterostructures. These structures were addressed using continuous wave photoexcitation only~\cite{Reshina2008}. Thereby the formation of EMP was established and the polaron energy was measured. Here we study the EMP dynamics using time-resolved photoluminescence (PL) as functions of magnetic field and temperature. The following Section ~\ref{sec:TB} presents the short theoretical background
of the exchange interaction in DMS and the EMP parameters. Section ~\ref{sec:ED} describes the investigated samples and the experimental technique. The experimental results and discussion are given in Sections ~\ref{sec:Spectroscopy} and \ref{sec:Dynamics} for time-integrated and time-resolved measurements, respectively. The discussion of the results is presented in section \ref{sec:Discussion}.

\section{Theoretical background}
\label{sec:TB}

In II-VI DMS the conduction band electrons with spin $s=1/2$ and the valence band holes with angular momentum $J=3/2$ are subject to the $sp-d$ exchange interaction with the localized spins of the Mn$^{2+}$ ions, which have angular momentum $S^{Mn}=5/2$. This results in giant Zeeman splitting of conduction and valence bands under applied magnetic field $\mathbf{B}$.  In the Faraday geometry, when the magnetic field $\mathbf{B}$ is oriented along the QW growth axis coinciding with the optical axis $\textbf{z}$, the heavy-hole exciton splits into two components with spin $-1$, which is composed of $s_z=+1/2$ and $J_z=-3/2$, and spin +1 constructed of $s_z=-1/2$ and $J_z=+3/2$. The energy splitting between these Zeeman sublevels is described by the well known relation \cite{Gaj79}
\begin{equation}
\label{eq:giantZeeman} \Delta E_z(B) = x N_0 (\alpha - \beta) \langle S_{z}^{Mn} (B) \rangle,
\end{equation}
where $x$ is the Mn$^{2+}$ concentration, $N_0 \alpha$ and $N_0 \beta$ are  the exchange constants for the conduction and valence bands, respectively, and $\langle S_{z}^{Mn} (B) \rangle$ is the thermal average of the Mn spin projection along ${\bf B}$. The latter is given by the modified Brillouin function B$_S$ for $S=5/2$:
\begin{equation}
\label{eq:Brillouin} \langle S_{z}^{Mn} (B) \rangle = S_{eff} \mathrm{B_{5/2}} \left[  \frac{5 \mu_B g_{Mn} B }{2 k_B(T+T_0)} \right],
\end{equation}
where $\mu_B$ is the Bohr magneton, $k_B$ is the Boltzmann constant, $T$ is the lattice (bath) temperature and $g_{Mn}=2.01$ is the Mn$^{2+}$ $g$ factor. $S_{eff}$ is the effective spin and $T_0$ is the effective temperature. These parameters permit a phenomenological description of the antiferromagnetic Mn-Mn exchange interaction. We use here the bulk values for the exchange integrals in Cd$_{1-x}$Mn$_{x}$Se:  $N_0\alpha=0.258$~eV and $N_0\beta=-1.238$~eV \cite{Dietl, ExchangeConst}. Typically the intrinsic Zeeman splittings of electron and hole and the electron-hole exchange interaction are much smaller compared with the effects induced by the $sp-d$ exchange interaction. Therefore, they are neglected.

The exciton magnetic polarons are formed due to polarization of the magnetic ion spins by the exchange field of the holes $B_{ex}$, which are dominant in the exchange interaction with the magnetic ions because the hole exchange integral is almost five times larger than that of the electrons. The resulting cloud of polarized Mn spins can be considered as magnetic molecule with a magnetic moment of hundreds of Bohr magnetons. The exchange field $B_{ex}$ is inversely proportional to the hole localization volume~\cite{Kavokin99}.

The EMP formation is accompanied by the decrease of the associated exciton energy by the polaron energy and can be treated as localization process. The primary localization of the exciton plays an important role in the EMP formation. Theory predicts that the stability conditions for free magnetic polarons can be hardly fulfilled in bulk DMS based on II-VI materials, where strong initial localization on alloy and/or magnetic fluctuations is required for EMP formation \cite{Ben93, Theory, Yakovlev10}. The situation changes with reduction of dimensionality, because the relatively weak primary localization of the exciton (e.g. on interface fluctuations in QWs) can be sufficient for further self-localization leading to a gain in exchange energy. Such self-localization implies a modification of the carrier wavefunctions in the EMP formation process, particulary squeezing of the hole wavefunction, which contributes to the polaron energy and affects the polaron dynamics and exciton radiative lifetime \cite{Number6,Kav93b,Dah95}.

The polaron energy $E_{MP}$ is related to the exchange field $B_{ex}$ by \cite{CPola_JETP95}
\begin{equation}
\label{eq:PolaronEnergy} E_{MP} = \frac{1}{2}\gamma B_{ex},
\end{equation}
where $\gamma=d \Delta E_z^{hh} / dB|_{B=0}$ is the slope of the heavy-hole giant Zeeman splitting measured in the Faraday geometry in the limit of weak magnetic fields. $\Delta E_z^{hh}(B)$ is connected with the exciton giant Zeeman splitting by the relation:
\begin{equation}
\label{eq:HoleZee} \Delta E_z^{hh}(B) = \frac{|\beta|}{|\alpha|+|\beta|}\Delta E_z(B).
\end{equation}
It is important to note here, that Eq.~(\ref{eq:PolaronEnergy}) is derived for localized magnetic polarons where the hole localization volume, and correspondingly the hole exchange field $B_{ex}$, do not change during EMP formation. This condition is fulfilled for the most studied (Cd,Mn)Te/(Cd,Mg)Te structures.

Experimentally, selective excitation of localized excitons gives the most direct and reliable access to the EMP parameters~\cite{Mackh94, Yakovlev10}. Thereby excitons are excited in the tail of localized states below the mobility edge, where further spectral diffusion along the tail of localized states is suppressed. Excitons can decrease their energy, however, by magnetic polaron formation and the energy shift of the emission line maximum $\hbar\omega_{PL}$ relative to the
excitation energy $\hbar\omega_{exc}$ is solely contributed by the polaron exchange shift $\Delta E$. The EMP formation time, $\tau_f$, typically is in the range of 50 to 200~ps~\cite{Linz94,Number6}. At times exceeding $\tau_f$ the polaron shift approaches the EMP energy
\begin{equation}
\label{eq:Form} \Delta E (t>>\tau_f) = E_{MP}.
\end{equation}
However, exciton recombination occurring with characteristic time $\tau_0$ may interrupt the polaron formation process. In this case the polaron shift observed in time-integrated PL spectra is smaller than $E_{MP}$. Therefore, time-resolved experiments, which allow one to measure both $\tau_f$ and $\tau_0$, are required for evaluation of the EMP energy.

An alternative technique for evaluating the EMP parameters is based on monitoring the magnetic-field-induced polarization of magnetic polarons, which reflects the orientation of the magnetic fluctuations inside the polaron volume, and therefore contains information about the hole localization volume and the hole exchange field~\cite{CPola_JETP95,Merkulov96,Kavokin99}. Time-resolved measurements are not required in this case. The circular polarization degree of the EMP emission, $\rho_c(B)$, is measured as function of the external magnetic field. Its slope in weak magnetic field $\theta=d\rho_c/dB|_{B=0}$ allows evaluation of the polaron energy
\begin{equation}
\label{eq:CPola}  E_{MP}=\frac{1}{2\pi k_B T}\left( \frac{\gamma}{\theta} \right)^2 .
\end{equation}
It was shown for (Cd,Mn)Te based samples that both techniques give very similar $E_{MP}$ values.\cite{CPola_JETP95,Kavokin99, Number6} Nevertheless, if the localization volume changes during the polaron formation process the theoretical approach should account for the self-localization effect.

\section{Experimental details}
\label{sec:ED}

The investigated (Cd,Mn)Se/(Cd,Mg)Se quantum well structures were grown by molecular-beam epitaxy on (001) oriented InAs substrates. Details of the growth technique were reported previously \cite{Reshina2006,Ivanov2004}. Most of the studies presented here were done on a multiple quantum well (MQW) structure (\#372) with five periods of Cd$_{0.935}$Mn$_{0.065}$Se/Cd$_{0.83}$Mg$_{0.17}$Se layers. The wells are composed of the DMS material with a width of 3.8 nm, sandwiched between nonmagnetic barriers with thicknesses of 9.7~nm. The barriers are sufficiently thick to decouple the electronic states in the neighboring QWs and, therefore, the QWs can be treated as isolated. On top of this structure a 0.1~$\mu$m-thick ${\rm Cd_{0.83}Mg_{0.17}Se}$ layer and a 5 nm-thick CdSe cap layer were grown.

Another studied CdSe/Cd$_{0.8}$Mg$_{0.2}$Se/Cd$_{0.927}$Mn$_{0.073}$Se double quantum well (DQW) structure (\#309) contains sequences of two QWs separated by a thin nonmagnetic barrier. The one, nonmagnetic CdSe QW has a width of 3.8~nm and the other, DMS Cd$_{0.927}$Mn$_{0.073}$Se QW has a width of 3.4~nm. A nonmagnetic Cd$_{0.8}$Mg$_{0.2}$Se barrier with thickness of 1.2~nm allows electronic coupling between the wells. The whole structure comprises 5 periods of these DQW layers whih are separated by 6.7 nm ${\rm Cd_{0.8}Mg_{0.2}Se}$ barriers. This sample has also a 0.1~$\mu$m-thick ${\rm Cd_{0.8}Mg_{0.2}Se}$ layer and a 5 nm-thick CdSe cap layer. In this study we use this sample as a reference for the properties of nonmagnetic CdSe QWs.

\begin{figure}
    \centering
    \includegraphics[width=\linewidth]{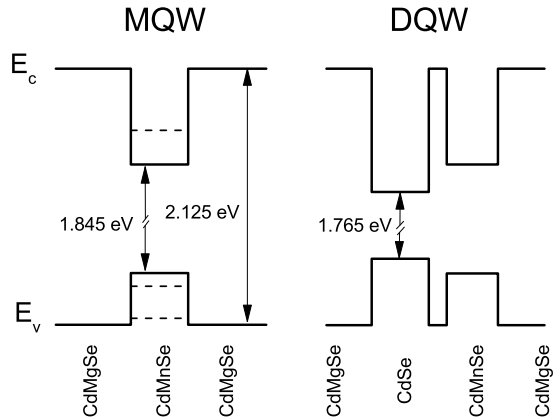}
    \caption{\label{fig:Structure} Band gap diagram of a single period of the Cd$_{0.935}$Mn$_{0.065}$Se/Cd$_{0.83}$Mg$_{0.17}$Se MQW sample and the CdSe/Cd$_{0.8}$Mg$_{0.2}$Se/Cd$_{0.927}$Mn$_{0.073}$Se DQW structure. }
    \label{fig1}
\end{figure}

The band gap diagrams of both structures are presented in Fig.~\ref{fig:Structure}. The band offsets in CdMnSe/CdMgSe are not yet well elaborated. Their optical properties were reported in Refs.~\onlinecite{Reshina2008} and \onlinecite{Reshina2006} where the valence band offset was estimated to be 38\% of the energy gap difference. This corresponds to potential barriers of about 180 and 100~meV for electrons and holes, respectively.

The samples were held in the variable temperature inset of a cryostat with a split-coil superconducting solenoid capable of generating magnetic fields $B\leq7$~T. They were either immersed in pumped liquid helium to reach a temperature $T=2$~K or were in contact with cold helium gas to measure temperature dependencies up to 30~K. The magnetic field was applied in Faraday geometry parallel to the structure growth axis ($\mathbf{B}\parallel\mathbf{z}$) and parallel to the light wave vector. The laser excitation was focused into a spot with a diameter of about 300~$\mu$m and with the excitation densities kept low enough to exclude heating effects.

For reflectivity measurements a halogen lamp was used as white light source. Photoluminescence (PL) was excited by a cw semiconductor diode laser with excitation energy at 2.54~eV, which exceeds the band gap of the Cd$_{0.83}$Mg$_{0.17}$Se  barriers (2.125~eV) and is referred to as above-barrier excitation. The circular polarization degree of the PL induced by an external magnetic field was measured as well. For that purpose the PL was detected in $\sigma^+$ and $\sigma^-$ polarization selected by rotation of a $\lambda/4$-wave plate in front of a Glan-Thomson prism placed before the installed monochromator. The degree of circular polarization is defined as $\rho_c = (I_+ - I_- )/(I_+ + I_-)$, where $I_+$ and $I_-$  are the PL intensities for $\sigma^+$ and $\sigma^-$ polarizations, respectively.

For time-resolved measurements we used a mode-locked Ti:Sapphire laser combined with a synchronously-pumped optical parametric oscillator equipped with an internal frequency-doubling unit for excitation. The system emitted pulses with a duration of about 1~ps and a spectral width of 1~nm at a repetition frequency of 76~MHz. The laser photon energy was tuned to $\hbar\omega_{exc} = 2.06$~eV in order to achieve below-barrier excitation of the investigated samples. Time-resolved measurements were also performed with above-barrier excitation $\hbar\omega_{exc} = 2.95$~eV. In this case the second harmonic unit was used directly after the Ti:Sapphire laser.

The photoluminescence signal was dispersed in a 0.5-meter monochromator with 6.33~nm/mm linear dispersion and was detected with a streak camera. The system time resolution was 20~ps. The EMP lifetimes were determined from the decay of the integral PL intensity, while the energy shift of the PL maxima allows one to trace the EMP formation dynamics~\cite{Number6, Yakovlev10}. Very similar results for the EMP dynamics were obtained for above- and below-barrier excitation. Also time-integrated PL spectra were measured with 1.35~nm/mm dispersion under pulsed photoexcitation using a charge-coupled-device detector connected to the same monochromator as the streak-camera.

\section{SPECTROSCOPY OF EXCITON MAGNETIC POLARON}
\label{sec:Spectroscopy}

\begin{figure}
\begin{minipage}{8.2cm}
 \epsfxsize=6.5 cm
 \centerline{\epsffile{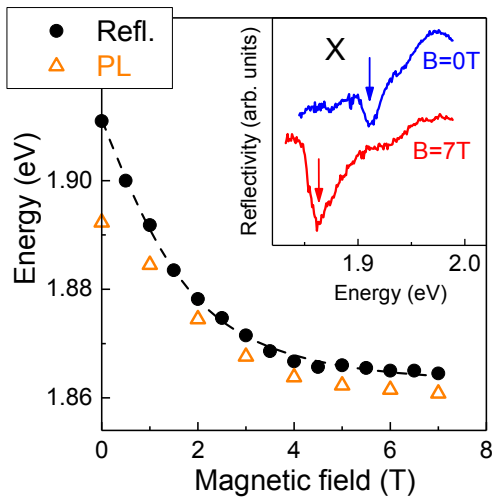}}
\caption{\label{fig:Zeeman} Magnetic field dependence of the exciton (X) energy in the MQW sample measured from reflectivity (circles) at $T = 2$~K. The dashed line is the fit with Eqs.~(\ref{eq:giantZeeman}, \ref{eq:Brillouin}). The peak energies of the time-integrated PL spectra measured under pulsed excitation with $\hbar\omega_{exc}=2.06$~eV is shown by the triangles. Inset: Reflectivity spectra for $B = 0$ and 7~T. The arrows indicate the exciton resonance position.}
\end{minipage}
\end{figure}

The giant Zeeman splitting of the exciton resonance in MQW sample was measured by means of reflectivity. Examples of reflectivity spectra at $B=0$ and 7~T are shown in the inset of Fig.~\ref{fig:Zeeman}. The heavy-hole exciton (X) resonance shows up as clear minimum, its energy position is marked by the arrow. In magnetic field the lower Zeeman sublevel can be clearly identified, while the upper one is broadened. The former one shifts to lower energies with increasing magnetic field by half of the total giant Zeeman splitting of the excitons, see closed circles in Fig.~\ref{fig:Zeeman}.  Fitting the shift data with Eqs.~(\ref{eq:giantZeeman},\ref{eq:Brillouin}), see the dashed line, we find $S_{eff}=0.98$ and $T_0 = 1.5$~K. In the limit of low magnetic fields we evaluate $d \Delta E_z/ dB|_{B=0}=42.8$~meV/T. This gives  $\gamma = 35$~meV/T, which we will use for evaluation of the EMP parameters.

\begin{figure}
\begin{minipage}{8.2cm}
 \epsfxsize=7.5 cm
 \centerline{\epsffile{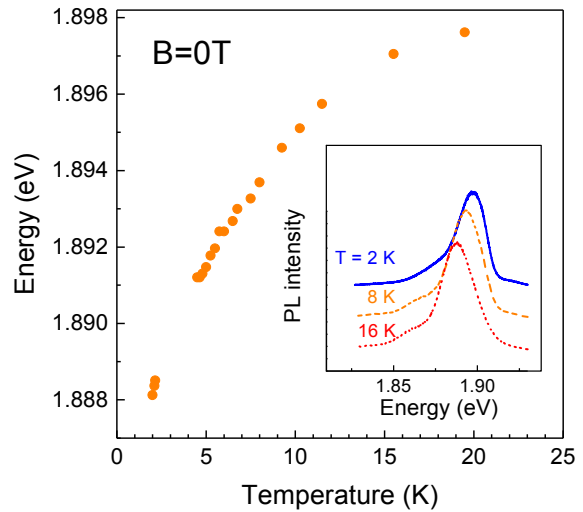}}
\caption{\label{fig:TdepPL} Temperature dependence of the PL peak maximum at $B = 0$~T. Inset: Time integrated PL spectra for $T = 2, 8$ and 16~K measured under cw excitation with $\hbar\omega_{exc}=2.54$~eV.}
\end{minipage}
\end{figure}

Additionally, Fig.~\ref{fig:Zeeman} shows the dependence of the PL maximum on magnetic field. An examplary PL spectrum at $B = 0$~T is presented in the inset of Fig.~\ref{fig:TdepPL}. The PL peak is red shifted at $T=2$~K by almost 20~meV with respect to the free exciton resonance deduced from the reflectivity spectrum (compare open and closed symbols in Fig.~\ref{fig:Zeeman}). Increasing the magnetic field leads to vanishing of the shift and narrowing of the PL line.

The temperature dependence of the PL maximum at $B=0$~T is shown in Fig.~\ref{fig:TdepPL}. Here a high-energy shift by 10~meV is observed with increasing temperature from 2 to 20~K. The temperature and magnetic field dependencies clearly evidence the EMP contribution to the PL spectra\cite{Mackh94, Yakovlev95}. Magnetic field application leads to suppression of the EMP and, therefore, the difference in the reflectivity and PL peak positions vanishes. The EMP can be also destroyed by a temperature increase due to
reduction of the magnetic susceptibility of the Mn$^{2+}$ ions.

\begin{figure}
\begin{minipage}{8.2cm}
 \epsfxsize=7 cm
 \centerline{\epsffile{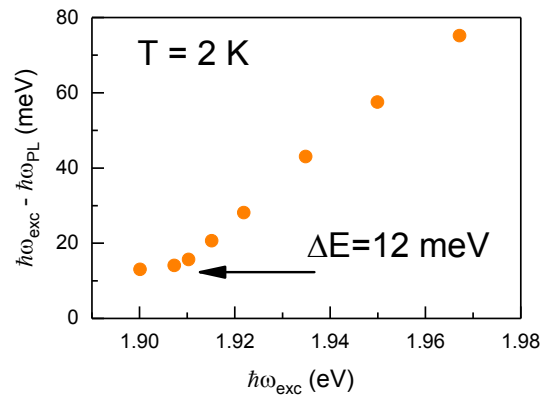}}
\caption{\label{fig:SelectiveExc} Energy separation between the PL peak maximum and the excitation energy as function of $\hbar\omega_{exc}$. Data were measured with pulsed excitation and time integrated detection.}
\end{minipage}
\end{figure}

\begin{figure}
\begin{minipage}{8.2cm}
 \epsfxsize=7 cm
 \centerline{\epsffile{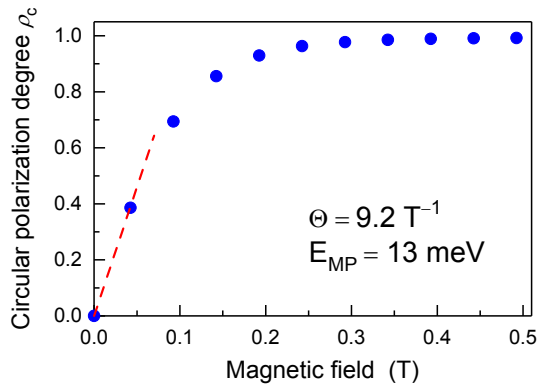}}
\caption{\label{fig:CPpola} Magnetic field dependence of the PL circular polarization degree $\rho_c(B)$. The line corresponds to a linear fit in weak magnetic fields. cw excitation with $\hbar\omega_{exc}=2.54$~eV, and time-integrated detection. $T = 2$~K.}
\end{minipage}
\end{figure}

First, we evaluate the polaron shift $\Delta E$ from time-integrated PL measurements under selective excitation of localized exciton states~\cite{Kusrayev-89,Mackh94,Yakovlev95,Kusrayev-96}. Figure~\ref{fig:SelectiveExc} gives the energy separation between the PL maximum $\hbar\omega_{PL}$ and the excitation laser energy
$\hbar\omega_{exc}$ as function of $\hbar\omega_{exc}$. The energy of 1.91~eV, where this dependence changes its character, can be associated with the exciton mobility edge within the band of localized states. Below this energy spectral diffusion of the exciton due to phonon-assisted tunneling does not occur during the exciton lifetime. Then the PL shift with respect to the excitation energy is determined by the magnetic polaron formation only and the PL peak starts to follow $\hbar\omega_{exc}$. The cutoff energy in the dependence corresponds to the polaron shift of $\Delta E = 12$~meV.

Second, we analyze the magnetic field dependence of the circular polarization degree due to exciton thermalization on the Zeeman sublevels. Figure \ref{fig:CPpola} shows this dependence of $\rho_c(B)$, which grows with $B$ before it saturates at a level of 0.99 already at $B=0.3$~T. In the range of weak magnetic fields we find $\theta =9.2$~T$^{-1}$. Using Eq.~(\ref{eq:CPola}) together with the value of $\gamma = 35$~meV/T evaluated from the reflectivity measurements we find $E_{MP} =13$~meV. This result is consistent with previous measurements~\cite{Reshina2008}. The EMP energy $E_{MP}$ determined with this method is comparable to the polaron shift $\Delta E$ measured under selective exciton excitation. This can be interpreted as fast EMP formation with $\tau_f < \tau_0$. However, a direct measurement of the polaron dynamics using time-resolved PL spectroscopy is required to assess this claim in detail.

\section{Dynamics of magnetic polaron}
\label{sec:Dynamics}

\begin{figure}
\begin{minipage}{\linewidth}
 \epsfxsize=8 cm
 \centerline{\epsffile{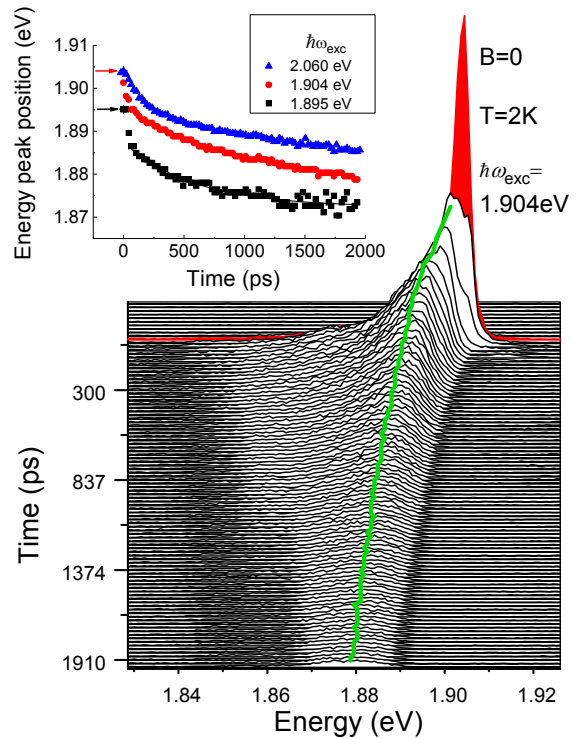}}
\caption{\label{fig:TRPLspectra} Time-resolved PL spectra at different time delays after quasi-resonant pulsed excitation with $\hbar\omega_{exc} =1.904$~eV. The red spectrum corresponds to $t=0$. The green line follows the PL maximum. Inset shows the temporal dependence of the PL maximum energy for different excitation energies. Arrows indicate the photon excitation energy. }
\end{minipage}
\end{figure}

\begin{figure}
\begin{minipage}{8.2cm}
 \epsfxsize=7 cm
 \centerline{\epsffile{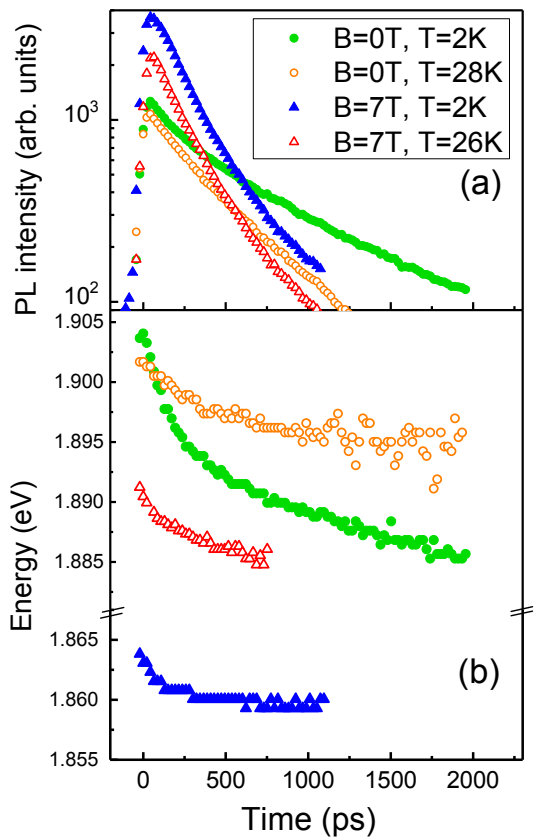}}
\caption{\label{fig:Transients} Temporal dependence of integral emission intensity (a) and PL peak position (b) for different magnetic fields and temperatures. $\hbar\omega_{exc} =2.06$~eV.}
\end{minipage}
\end{figure}

Temporally-resolved and spectrally-resolved PL spectra measured with the streak-camera allow us to study the EMP formation dynamics. PL spectra at different delay times $t$ after the excitation pulse are shown in Fig.~\ref{fig:TRPLspectra} at zero magnetic field. The PL line shifts continuously to lower energies with increasing time indicating the non-magnetic exciton localization and magnetic polaron formation. The polaron formation can be treated as magnetic localization of excitons. This shift is maintained also under resonant excitation of the exciton with $\hbar\omega_{exc} = 1.895$~eV as seen from the inset of Fig.~\ref{fig:TRPLspectra}. Spectrally integrated PL transients $I(t)$ and the PL peak energy position $E(t)$ at $B=0$ and 7~T for low and high temperatures are shown in Figs.~\ref{fig:Transients}(a) and \ref{fig:Transients}(b), respectively. As mentioned above, at $B = 0$~T and $T =2$~K we observe a strong low energy shift relative to the initial value of $E(t=0) = 1.904$~eV to $E(t\rightarrow \infty ) = 1.886$~eV at the longest measured delay of $t = 2$~ns. Application of magnetic field or increase of temperature leads to a significant suppression of this energy shift, which we attribute to reduction of the EMP energy. Interestingly, we observe a significant shortening of the exciton population decay with increase of $B$ or $T$ [see Fig.~\ref{fig:Transients}(a)].

\begin{figure}
    \centering
    \includegraphics[width=\linewidth]{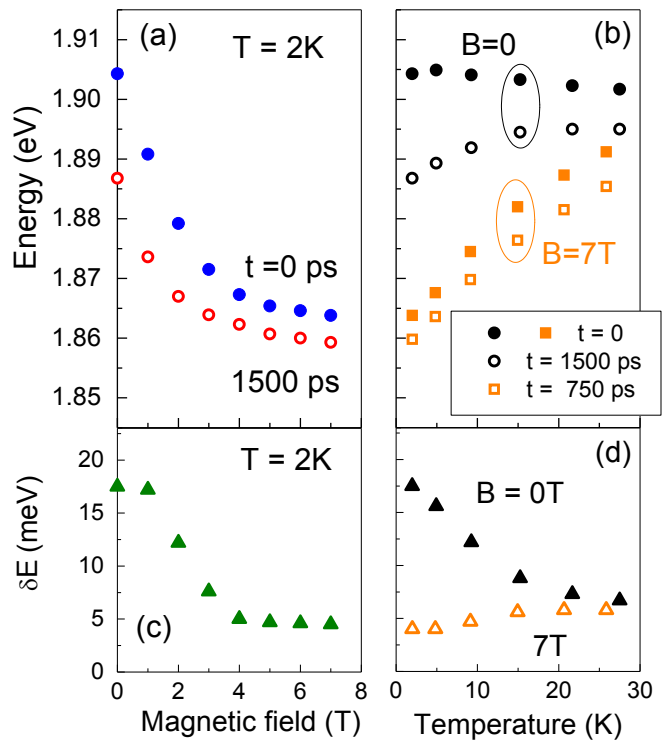}
\caption{\label{fig:Energies} Magnetic field (a) and temperature (b) dependencies of the PL peak positions at zero ($t=0$~ps) and maximum ($t = 1500$ or 750~ps) delay times after the excitation pulse. Magnetic field  (c) and temperature (d) dependencies of the energy difference between the values at minimum and maximum delay. $\hbar\omega_{exc} =2.06$~eV.}
\end{figure}

In order to quantify the results, we evaluate the energy positions $E(0)$ and $E(\infty)$ as function of magnetic field and temperature. The data are plotted in Figs.~\ref{fig:Energies}(a) and \ref{fig:Energies}(b). The final energy $E(\infty)$ is taken at the longest delay time $t$ where the position of the PL peak could be still reliably determined. Additionally, the dependencies of the energy difference  $\delta E = E(0)-E(\infty)$ on magnetic field and temperature are shown in Figs.~\ref{fig:Energies}(c) and \ref{fig:Energies}(d).
The magnetic field dependence of $E(0)$ follows closely the data for the lowest exciton resonance deduced from the reflectivity measurements (see Fig.~\ref{fig:Zeeman}), as expected if the laser excitation pulses photogenerate free excitons. Due to EMP suppression the energy difference $\delta E$ decreases from 18 to 4~meV with increasing magnetic field up to $B>4$~T, above which it stays constant, see Fig.~\ref{fig:Energies}(c). This behavior indicates the existence of additional localization, which does not vanish in magnetic field and, therefore, has a non-magnetic origin.

The existence of non-magnetic localization under non-selective exciton excitation is quite typical. Often it is related with exciton localization on composition fluctuations in ternary alloys as well as interface fluctuations in QWs. Thus we conclude, that at $B>4$~T the EMP is almost fully suppressed while the $\delta E=4$~meV shift corresponds to non-magnetic localization with a characteristic energy $E_{NM}$. The temperature dependence at $B=7$~T in Fig.~\ref{fig:Energies}(d) shows that the non-magnetic localization contribution slightly increases from 4~meV at $T = 2$~K to 6~meV at 26~K. Simultaneously, at $B = 0$~T the energy shift $\delta E$ decreases with temperature and finally tends to the value of $E_{NM}$ at $T = 28$~K. Such behavior is expected since the EMP energy decreases for a lower magnetic susceptibility of the Mn$^{2+}$ spin system.

Further insight in the magnetic polaron dynamics is obtained from the magnetic field and temperature dependencies of the time constants $\tau_0$ and $\tau_E$ characterizing the decay of the exciton population $I(t)$ and the energy shift $\delta E(t)$, respectively. Note that at $B = 0$~T and $T = 2$~K, where the polaron energy shift is most pronounced, the intensity transient and the energy peak position follow a non-exponential decay. Therefore $\tau_0$ and $\tau_E$ are determined as the moments when $I(t)$ and $\delta E(t)$ are reduced by the factor of $e=2.72$.

\begin{figure}
\begin{minipage}{8.2cm}
 \epsfxsize=7 cm
 \centerline{\epsffile{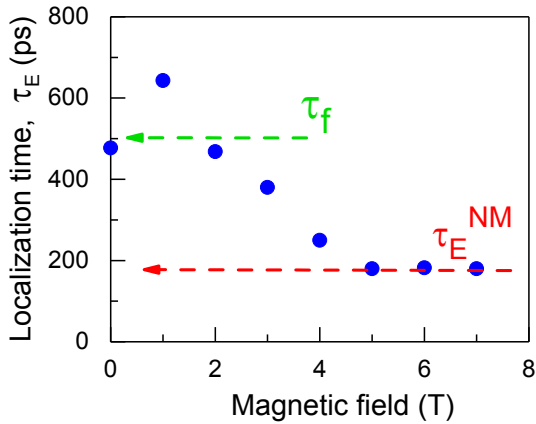}}
\caption{\label{fig:FormTimes}  Magnetic field dependence of the localization time $\tau_E$. $\hbar\omega_{exc}=2.06$~eV. Arrows indicate the magnetic polaron formation time $\tau_f$ and the non-magnetic exciton localization time $\tau_E^{NM}.$
}
\end{minipage}
\end{figure}

First, we focus on $\tau_E$, whose magnetic field dependence is presented in Fig.~\ref{fig:FormTimes}. In strong magnetic fields the EMP is suppressed and only non-magnetic localization occurs on a time scale of $\tau_E^{NM}=180$~ps. At zero magnetic field both non-magnetic and magnetic localization are present producing the total shift of $\delta E = E_{NM} + E_{MP} = 18$~meV. The timescale of this localization at $T=2$~K is given by $\tau_E = 480$~ps, which is slightly shorter than the exciton recombination time $\tau_0$. Assuming that the non-magnetic localization is independent of $B$ we put $E_{NM} = 4$~meV and evaluate the EMP energy $E_{MP} = 14$~meV. Here, the nonmagnetic contribution is significantly smaller than the EMP energy and therefore we attribute the localization time at $B=0$ to the formation time of the MP $\tau_f\approx500$~ps. Such long EMP formation time is quite a surprising result compared to previous studies for (Cd,Mn)Te and (Zn,Mn)Se based heterostructures for which times in the order of $50-200$~ps were reported \cite{Linz94, Number6, Henneberger96}.

\begin{figure}
\begin{minipage}{8.2cm}
 \epsfxsize=8 cm
 \centerline{\epsffile{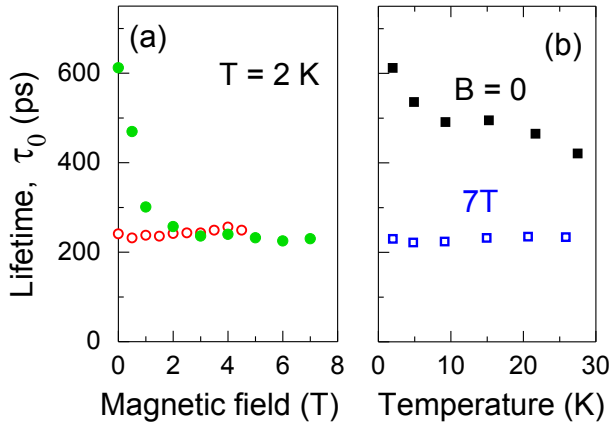}}
\caption{\label{fig:Times} Magnetic field (a) and temperature (b) dependencies of the exciton lifetime. Open circles in (a) correspond to the lifetime of the exciton in the non-magnetic CdSe well of the DQW structure. $\hbar\omega_{exc}=2.06$~eV.}
\end{minipage}
\end{figure}

The dependencies of the PL decay time on $B$ and $T$ are shown in Fig.~\ref{fig:Times}. We observe decreases of $\tau_0$ from 600~ps to either 200~ps or 400~ps with increase of magnetic field or temperature, respectively. Additionally, in Fig.~\ref{fig:Times}(a) we present the data for the $B$-dependence of $\tau_0$ in the non-magnetic well of the DQW structure. In this case $\tau_0=250$~ps is independent of magnetic field strength as expected, i.e. we obtain the same recombination time as in the magnetic QW, when the EMP is suppressed.

\section{Discussion}
\label{sec:Discussion}

A slow EMP dynamics is expected in case of self-localization which may take place if the initial localization on magnetic fluctuations is relatively weak. During EMP formation the exciton localization is enhanced leading to an increase of $E_{MP}$. The theoretical consideration in Ref.~\onlinecite{Number6} shows that the polaron self-localization in a two-dimensional system increases not only the polaron energy, but also the formation time
\begin{equation}
\label{eq:FormationTime}  \tau_{f} = \tau_s \frac{E_{MP}}{\Delta E_{eq}(0)},
\end{equation}
where $\tau_s$ is the EMP formation time without self-localization and $\Delta E_{eq}(0)$ is the equilibrium EMP energy at $t=0$, i.e. without self-localization. Since $\Delta E_{eq}(0) < E_{MP}$, the formation time becomes longer than $\tau_s$.

Previous studies reported no strong deviations of $\tau_f$ from $\tau_s$, except for (Cd,Mn)Te QW quantum wells, where for samples with low Mn$^{2+}$
concentration ($x=0.1$) a two-fold increase of the formation time was reported \cite{Linz94}. The intrinsic formation time $\tau_s$ in (Cd,Mn)Te QW structures was measured to be in the range of
150~ps. Similar times are expected for (Cd,Mn)Se. However, we find that in the studied (Cd,Mn)Se structures $\tau_f=500$~ps which is more than 3 times longer than $\tau_s$. Obviously, in these structures exciton self-localization has to play an important role in the EMP formation.

The equilibrium EMP energy at the initial moment of time $\Delta E_{eq}(0) = 4$~meV can be evaluated from Eq.~\eqref{eq:FormationTime} using $\tau_f=500$~ps, $\tau_s=150$~ps and $E_{MP}=14$~meV. The considerable self-localization of EMP in (Cd,Mn)Se-based QWs can be related to different factors: First, the smaller heavy-hole effective mass of only $m_{hh}^*=$0.45 in CdSe as compared to 0.8 in CdTe, which leads to
weaker confinement in the magnetic quantum well and a larger penetration of the wavefunction into the non-magnetic barriers. Second, the relatively small concentration of Mn$^{2+}$ ions. Both factors may lead to weaker initial magnetic localization in the studied structure, which is critical for self-localization.

The polaron shift $\Delta E=12$~meV evaluated from time-integrated measurements under resonant exciton excitation is in good agreement with the time-resolved data yielding $E_{MP}=14$~meV. In fact, no large difference between these values is expected because the polaron formation time, although being long, does not exceed the exciton radiative lifetime of 600~ps. We note, however, that the small value of $\Delta E_{eq}(0) = 4$~meV does not coincide with the polaron energy of 13~meV extracted from the magnetic
field dependence of the PL circular polarization degree. The latter reflects the initial distribution of magnetic fluctuations and, therefore, should correspond to $\Delta E_{eq}(0)$. This fact deserves attention and the dynamics of magnetic polaron in weak magnetic fields should be studied more detailed in that respect in the future.

The self-localization process is accompanied by a shrinkage of the heavy-hole envelope wavefunction. As a result the overlap of the electron and the hole wave functions changes, leading to a deceleration or acceleration of the PL intensity decay. Indeed we observe a non-exponential decay of $I(t)$ with a rather long radiative lifetime of 600~ps at $B=0$~T and low temperatures, where the polaron localization is most pronounced (see Fig.~\ref{fig:Transients}). Moreover, the radiative decay becomes significantly shorter, when the EMP is suppressed by application of magnetic fields $B\geq$4~T and/or at elevated temperatures above 20~K. The exciton lifetime $\tau_0$ drops to 200~ps at $B\geq$4~T. For an elevated temperature of 28~K we find $\tau_0=400$~ps.

It is important to distinguish between out of plane and lateral squeezing of the heavy hole wavefunction in the QW during the process of EMP formation. The first scenario takes place if the hole confinement in the QW is weak. Then, in narrow QWs, e.g. of 3.8~nm as studied experimentally in this paper, the electron wavefunction is strongly localized in the DMS QW, while the hole wavefunction has considerable penetration in the nonmagnetic barriers. That would result in: (i) a longer exciton recombination, due to a smaller overlap of the electron and hole wavefunctions, and (ii) a specific dynamics of the EMP in whose formation the self-localization process of the hole wavefunction should contribute considerably. Both features are in line with our experiment. However, we exclude this scenario because it should also lead to a nonlinear dependence of the Zeeman splitting in small magnetic fields, which is in contrast to the data presented in Fig.~\ref{fig:Zeeman}. Moreover, previous estimations of the valence band offset show that the QW potential in the studied structures is about 100~meV, which is significantly larger than $E_{MP}$. The second scenario of in-plane self-localization is realized if electron and hole become separated spatially in the QW plane due to different origins of their localization potentials. The electrons are localized on non-magnetic potential fluctuations, while the holes with strong exchange interaction with the magnetic ions are localized on the magnetic fluctuations.  Due to self-localization of the EMP the in-plane squeezing of the heavy hole wavefunction reduces the overlap with the electron wavefunction and consequently the exciton lifetime increases. This second scenario is most relevant for our case.

We emphasize that the strong correlation between the radiative lifetime and the magnetic polaron formation supports the conclusion about the presence of self-localization. Note that previous observations of self-localization were based on the comparison of magnetic QWs with different well widths \cite{Linz94}. Here we show directly how EMP suppression by magnetic field or temperature can change the exciton dynamics.

\section{Conclusions}
\label{sec:Conclusions}

We have studied time-resolved PL spectra of (Cd,Mn)Se/(Cd,Mg)Se quantum wells in magnetic fields up to 7~T applied in the Faraday geometry and in the temperature range between 2 and 30~K. The formation of exciton magnetic polarons has been established using different spectroscopic techniques. We distinguish between magnetic and non-magnetic localization in the formation process. The magnetic polaron energy at zero magnetic field extracted from time-resolved data gives amounts to 14~meV, while non-magnetic contribution at low temperatures of 2~K is around 4~meV. The magnetic polaron formation time of 500~ps is comparable with the exciton recombination time of 600~ps. We associate the long formation time to strong self-localization, accompanied with squeezing of the heavy-hole envelope wavefunction. This conclusion is supported by the decrease of the exciton lifetime down to 200~ps in strong magnetic fields and to 400~ps at elevated temperatures.

\section{Acknowledgements}
\label{sec:acknowledgements}
We acknowledge the financial support of the Russian Science Foundation (Grant No. 14-42-00015) and the Deutsche Forschungsgemeinschaft in the frame of ICRC TRR 160. The MBE growth studies at Ioffe Institute were supported by RFBR Grant No. 15-52-12014.

\end{document}